\documentclass[prl,twocolumn,showpacs,superscriptaddress,floatfix]{revtex4}
\usepackage{epsfig}
\usepackage{graphicx}
\usepackage{dcolumn}

\def\be{\begin{equation}}
\def\ee{\end{equation}}
\def\etm{E_{T}}
\def\ptm{p_{T}}

\begin{document}

\pagestyle{empty}






\title{Search for a High-Mass Diphoton State and Limits on Randall-Sundrum Gravitons at CDF}
\affiliation{Institute of Physics, Academia Sinica, Taipei, Taiwan 11529, Republic of China} 
\affiliation{Argonne National Laboratory, Argonne, Illinois 60439} 
\affiliation{Institut de Fisica d'Altes Energies, Universitat Autonoma de Barcelona, E-08193, Bellaterra (Barcelona), Spain} 
\affiliation{Baylor University, Waco, Texas  76798} 
\affiliation{Istituto Nazionale di Fisica Nucleare, University of Bologna, I-40127 Bologna, Italy} 
\affiliation{Brandeis University, Waltham, Massachusetts 02254} 
\affiliation{University of California, Davis, Davis, California  95616} 
\affiliation{University of California, Los Angeles, Los Angeles, California  90024} 
\affiliation{University of California, San Diego, La Jolla, California  92093} 
\affiliation{University of California, Santa Barbara, Santa Barbara, California 93106} 
\affiliation{Instituto de Fisica de Cantabria, CSIC-University of Cantabria, 39005 Santander, Spain} 
\affiliation{Carnegie Mellon University, Pittsburgh, PA  15213} 
\affiliation{Enrico Fermi Institute, University of Chicago, Chicago, Illinois 60637} 
\affiliation{Comenius University, 842 48 Bratislava, Slovakia; Institute of Experimental Physics, 040 01 Kosice, Slovakia} 
\affiliation{Joint Institute for Nuclear Research, RU-141980 Dubna, Russia} 
\affiliation{Duke University, Durham, North Carolina  27708} 
\affiliation{Fermi National Accelerator Laboratory, Batavia, Illinois 60510} 
\affiliation{University of Florida, Gainesville, Florida  32611} 
\affiliation{Laboratori Nazionali di Frascati, Istituto Nazionale di Fisica Nucleare, I-00044 Frascati, Italy} 
\affiliation{University of Geneva, CH-1211 Geneva 4, Switzerland} 
\affiliation{Glasgow University, Glasgow G12 8QQ, United Kingdom} 
\affiliation{Harvard University, Cambridge, Massachusetts 02138} 
\affiliation{Division of High Energy Physics, Department of Physics, University of Helsinki and Helsinki Institute of Physics, FIN-00014, Helsinki, Finland} 
\affiliation{University of Illinois, Urbana, Illinois 61801} 
\affiliation{The Johns Hopkins University, Baltimore, Maryland 21218} 
\affiliation{Institut f\"{u}r Experimentelle Kernphysik, Universit\"{a}t Karlsruhe, 76128 Karlsruhe, Germany} 
\affiliation{High Energy Accelerator Research Organization (KEK), Tsukuba, Ibaraki 305, Japan} 
\affiliation{Center for High Energy Physics: Kyungpook National University, Taegu 702-701, Korea; Seoul National University, Seoul 151-742, Korea; SungKyunKwan University, Suwon 440-746, Korea} 
\affiliation{Ernest Orlando Lawrence Berkeley National Laboratory, Berkeley, California 94720} 
\affiliation{University of Liverpool, Liverpool L69 7ZE, United Kingdom} 
\affiliation{University College London, London WC1E 6BT, United Kingdom} 
\affiliation{Centro de Investigaciones Energeticas Medioambientales y Tecnologicas, E-28040 Madrid, Spain} 
\affiliation{Massachusetts Institute of Technology, Cambridge, Massachusetts  02139} 
\affiliation{Institute of Particle Physics: McGill University, Montr\'{e}al, Canada H3A~2T8; and University of Toronto, Toronto, Canada M5S~1A7} 
\affiliation{University of Michigan, Ann Arbor, Michigan 48109} 
\affiliation{Michigan State University, East Lansing, Michigan  48824} 
\affiliation{University of New Mexico, Albuquerque, New Mexico 87131} 
\affiliation{Northwestern University, Evanston, Illinois  60208} 
\affiliation{The Ohio State University, Columbus, Ohio  43210} 
\affiliation{Okayama University, Okayama 700-8530, Japan} 
\affiliation{Osaka City University, Osaka 588, Japan} 
\affiliation{University of Oxford, Oxford OX1 3RH, United Kingdom} 
\affiliation{University of Padova, Istituto Nazionale di Fisica Nucleare, Sezione di Padova-Trento, I-35131 Padova, Italy} 
\affiliation{LPNHE, Universite Pierre et Marie Curie/IN2P3-CNRS, UMR7585, Paris, F-75252 France} 
\affiliation{University of Pennsylvania, Philadelphia, Pennsylvania 19104} 
\affiliation{Istituto Nazionale di Fisica Nucleare Pisa, Universities of Pisa, Siena and Scuola Normale Superiore, I-56127 Pisa, Italy} 
\affiliation{University of Pittsburgh, Pittsburgh, Pennsylvania 15260} 
\affiliation{Purdue University, West Lafayette, Indiana 47907} 
\affiliation{University of Rochester, Rochester, New York 14627} 
\affiliation{The Rockefeller University, New York, New York 10021} 
\affiliation{Istituto Nazionale di Fisica Nucleare, Sezione di Roma 1, University of Rome ``La Sapienza," I-00185 Roma, Italy} 
\affiliation{Rutgers University, Piscataway, New Jersey 08855} 
\affiliation{Texas A\&M University, College Station, Texas 77843} 
\affiliation{Istituto Nazionale di Fisica Nucleare, University of Trieste/\ Udine, Italy} 
\affiliation{University of Tsukuba, Tsukuba, Ibaraki 305, Japan} 
\affiliation{Tufts University, Medford, Massachusetts 02155} 
\affiliation{Waseda University, Tokyo 169, Japan} 
\affiliation{Wayne State University, Detroit, Michigan  48201} 
\affiliation{University of Wisconsin, Madison, Wisconsin 53706} 
\affiliation{Yale University, New Haven, Connecticut 06520} 
\author{T.~Aaltonen}
\affiliation{Division of High Energy Physics, Department of Physics, University of Helsinki and Helsinki Institute of Physics, FIN-00014, Helsinki, Finland}
\author{A.~Abulencia}
\affiliation{University of Illinois, Urbana, Illinois 61801}
\author{J.~Adelman}
\affiliation{Enrico Fermi Institute, University of Chicago, Chicago, Illinois 60637}
\author{T.~Affolder}
\affiliation{University of California, Santa Barbara, Santa Barbara, California 93106}
\author{T.~Akimoto}
\affiliation{University of Tsukuba, Tsukuba, Ibaraki 305, Japan}
\author{M.G.~Albrow}
\affiliation{Fermi National Accelerator Laboratory, Batavia, Illinois 60510}
\author{S.~Amerio}
\affiliation{University of Padova, Istituto Nazionale di Fisica Nucleare, Sezione di Padova-Trento, I-35131 Padova, Italy}
\author{D.~Amidei}
\affiliation{University of Michigan, Ann Arbor, Michigan 48109}
\author{A.~Anastassov}
\affiliation{Rutgers University, Piscataway, New Jersey 08855}
\author{K.~Anikeev}
\affiliation{Fermi National Accelerator Laboratory, Batavia, Illinois 60510}
% ...
\collaboration{CDF Collaboration\footnote{With visitors from $^a$University of Athens, 15784 Athens, Greece, 
$^b$University of Bristol, Bristol BS8 1TL, United Kingdom, 
$^c$University Libre de Bruxelles, B-1050 Brussels, Belgium, 
$^d$Cornell University, Ithaca, NY  14853, 
$^e$University of Cyprus, Nicosia CY-1678, Cyprus, 
$^f$University College Dublin, Dublin 4, Ireland, 
$^g$University of Edinburgh, Edinburgh EH9 3JZ, United Kingdom, 
$^h$University of Heidelberg, D-69120 Heidelberg, Germany, 
$^i$Universidad Iberoamericana, Mexico D.F., Mexico, 
$^j$University of Manchester, Manchester M13 9PL, England, 
$^k$Nagasaki Institute of Applied Science, Nagasaki, Japan, 
$^l$University de Oviedo, E-33007 Oviedo, Spain, 
$^m$University of London, Queen Mary College, London, E1 4NS, England, 
$^n$University of California Santa Cruz, Santa Cruz, CA  95064, 
$^o$Texas Tech University, Lubbock, TX  79409, 
$^p$University of California, Irvine, Irvine, CA  92697, 
$^q$IFIC(CSIC-Universitat de Valencia), 46071 Valencia, Spain, 
}}
\noaffiliation

\begin{abstract}

We have performed a search for new particles which decay 
to two photons using 1.2~fb$^{-1}$ of integrated luminosity from $p\bar p$ collisions 
at $\sqrt{s}$ = 1.96 TeV collected using the
CDF II Detector at the Fermilab Tevatron. 
We find the diphoton mass spectrum to be in agreement
with the standard model expectation, and set
limits on the cross section times
branching ratio 
for the Randall-Sundrum graviton, as a
function of diphoton mass. 
We subsequently derive
lower limits for the graviton mass of 
230~GeV$/c^{2}$ and 850~GeV$/c^{2}$, at the 95\% confidence level,
for coupling parameters 
($k/\overline{M}_{Pl}$) 
of 0.01 and 0.1 respectively.  
\end{abstract}




\maketitle


\clearpage

\par A potential signature for new, heavy particles is a narrow mass
resonance decaying to two energetic photons. Such a
signature could arise from extra spatial dimensions, as in
the Randall-Sundrum (RS) model~\cite{RS}.
In this Letter we present the results of the search for this 
signature in $p\bar{p}$ collision data from the CDF II
Detector at the Fermilab Tevatron.

Some string theories propose that 
as many as seven new spatial dimensions may exist, with their 
geometry potentially responsible for the apparent weakness of gravity (the
hierarchy problem). 
In the minimal RS model, 
this geometry is a five-dimensional space with negative curvature, bounded by
two three-dimensional, spatially extended membranes (or `branes'). 
These branes are separated by a distance 
$\pi r_c$, where  $r_c$ is the compactification radius. 
Using $\phi$ as the coordinate of the extra dimension 
($0\leq \phi \leq \pi$), 
the standard model (SM) particles 
are confined to the ``TeV'' brane,
located at $\phi = \pi$, while the gravitational wavefunction is localized at
$\phi = 0$. The scale of physical phenomena on the TeV brane is specified by
$\Lambda_\pi = \overline{M}_{Pl}~e^{-k r_c \pi}$, where 
$\overline{M}_{Pl} = M_{Pl}/ \sqrt{8 \pi} = 2.4\times 10^{18}$~GeV 
is the effective four-dimensional (reduced)
Planck scale and $k$ is a curvature parameter of the order of the Planck scale. 

To remove the hierarchy between gravity and the electroweak force, 
$\Lambda_\pi$ is set to approximately 1~TeV. The Planck scale 
then arises from the small overlap of the graviton wave function with
the TeV brane in the fifth dimension, and the compactification scale
gives rise to a Kaluza Klein tower of graviton states, an infinite set of 
four-dimensional particles with increasing masses.
This mass spectrum is given by 
$m_n = x_n (k/\overline{M}_{Pl}) \Lambda_\pi$, where $x_n$ are the roots
of the first-order Bessel function, and the states couple with strength 
$1/\Lambda_\pi$. Thus, the
widths and masses of the resonances are dependent on the parameter
$k/\overline{M}_{Pl}$.

The values of $k$ must be large enough to be consistent with
the apparent weakness 
of gravity, but small enough to prevent the theory from becoming 
non-perturbative~\cite{BF}. 
Given these considerations, we examine values in the range
$0.01 < k/\overline{M}_{Pl} < 0.1$.
For this range, graviton production results in a diphoton mass peak 
narrower than the CDF detector resolution. 
The spin-2 nature of the graviton, decaying by either $s$- 
or $p$-wave states, favors searches in the 
diphoton channel, where the branching ratio is twice 
that of any single dilepton channel. 

\par Existing lower mass limits on RS gravitons 
from a search using the D\O~detector,
in the combined diphoton, dielectron, and dimuon channels, 
are 250~GeV$/c^{2}$ for $k/\overline{M}_{Pl}=0.01$ and 
785~GeV$/c^{2}$ for $k/\overline{M}_{Pl}=0.1$ 
at the 95\% confidence level (C.L.)~\cite{D0PrevBestRSLimits}. 
The previous limits from the CDF collaboration are from a combined 
dielectron and dimuon search, with limits of 170~GeV$/c^{2}$ and 710~GeV$/c^{2}$ for 
$k/\overline{M}_{Pl}=0.01$ and 0.1 
respectively~\cite{CDFPrevBestRSLimits}.

This search uses 1.2~$\rm{fb^{-1}}$ of integrated luminosity 
collected by the CDF II Detector operating at $\sqrt{s} = 1.96$~TeV. 
The detector~\cite{CDF} is approximately 
forward-backward and azimuthally symmetric.
A seven-layer (eight-layer in the forward region) 
silicon tracker~\cite{SiliconTracker} is surrounded  
by an open-cell drift chamber (COT)~\cite{COT}. 
The fiducial coverage of the COT is $|\eta|<1.0$, and the silicon detector 
extends the tracking coverage to $|\eta|<2.0$. The
integrated tracking system is contained within a superconducting solenoid,
providing a 1.4~T magnetic field. 
Surrounding these are the electromagnetic 
and hadronic calorimeters~\cite{Calorimeter}, divided into ``central'' 
($|\eta|< 1.1$) and ``plug'' ($1.1<|\eta|<3.6$) regions, providing 
measurements of both shower energy and position. 
At the approximate electromagnetic shower maximum, the calorimeters 
contain fine-grained detectors~\cite{ShowerMax} that measure the 
shower shape and centroid position in the two dimensions
transverse to the shower development.
Surrounding these detectors is a system of 
muon detectors~\cite{Muon}. Three levels of 
real-time trigger systems are used to filter events.

\par The collision events recorded for analysis were selected by at 
least one of four triggers. Two of the triggers require 
two clusters of electromagnetic energy~\cite{CDF}: one of these triggers requires 
both clusters to have transverse energy $\etm>12$~GeV, and to be isolated 
in the calorimeter; 
while the other requires the two clusters to have $\etm>18$~GeV, 
but makes no isolation requirement. 
To ensure very high trigger efficiency at the largest photon $\etm$,
events are also accepted from two single-photon triggers.
The first of these triggers requires $\etm>50$~GeV without 
an isolation criterion; 
the second requires $\etm>70$~GeV, without an isolation ctriteria, 
and relaxed requirements on the hadronic energy 
associated with cluster.
The combination of these triggers is effectively 100\% efficient for 
the kinematic region used in this search for diphoton events 
above a mass of 50~GeV$/c^2$.

The triggered events are required to have been recorded 
while the relevant detector elements were fully operational.
Events are examined in 
which either both photons are in the central calorimeter 
(approximately in the region $|\eta|< 1.04$), or one photon is 
in the central and the other is in the plug calorimeter 
(over the region $1.2<|\eta|<2.8$). The data were collected between 
February, 2002 and February, 2006 and correspond to an integrated luminosity of 
1.2~$\rm{fb^{-1}}$ for the central-central (CC) diphoton sample 
and 1.1~$\rm{fb^{-1}}$ for the central-plug (CP) sample. 
The luminosity of the CP sample is reduced because good 
silicon detector conditions are required for tracking in the plug region.

In both the CC and CP cases, each of the two photons is required to produce a highly
electromagnetic energy 
cluster with $\etm>15$~GeV and together to have a 
reconstructed diphoton invariant mass 
greater than 30~GeV$/c^{2}$. 
Both are required to be in the fiducial region of the 
shower maximum detectors and to pass the following 
photon identification requirements:
transverse shower profiles consistent with a single photon, 
additional transverse energy in the calorimeter in a cone of angular 
radius R = $\sqrt{(\Delta \phi)^{2} + (\Delta \eta)^{2}} = 0.4$ 
around the photon candidate $<$ 2~GeV, 
and the scalar sum of the transverse momentum $\ptm$ of the tracks in the same 
cone $<$ 2~GeV$/c$. 
Photons in the central detector
are required to produce isolated energy clusters in the shower maximum detector.

\begin{figure}[h]
\begin{center}
\includegraphics[width=1.0\linewidth]{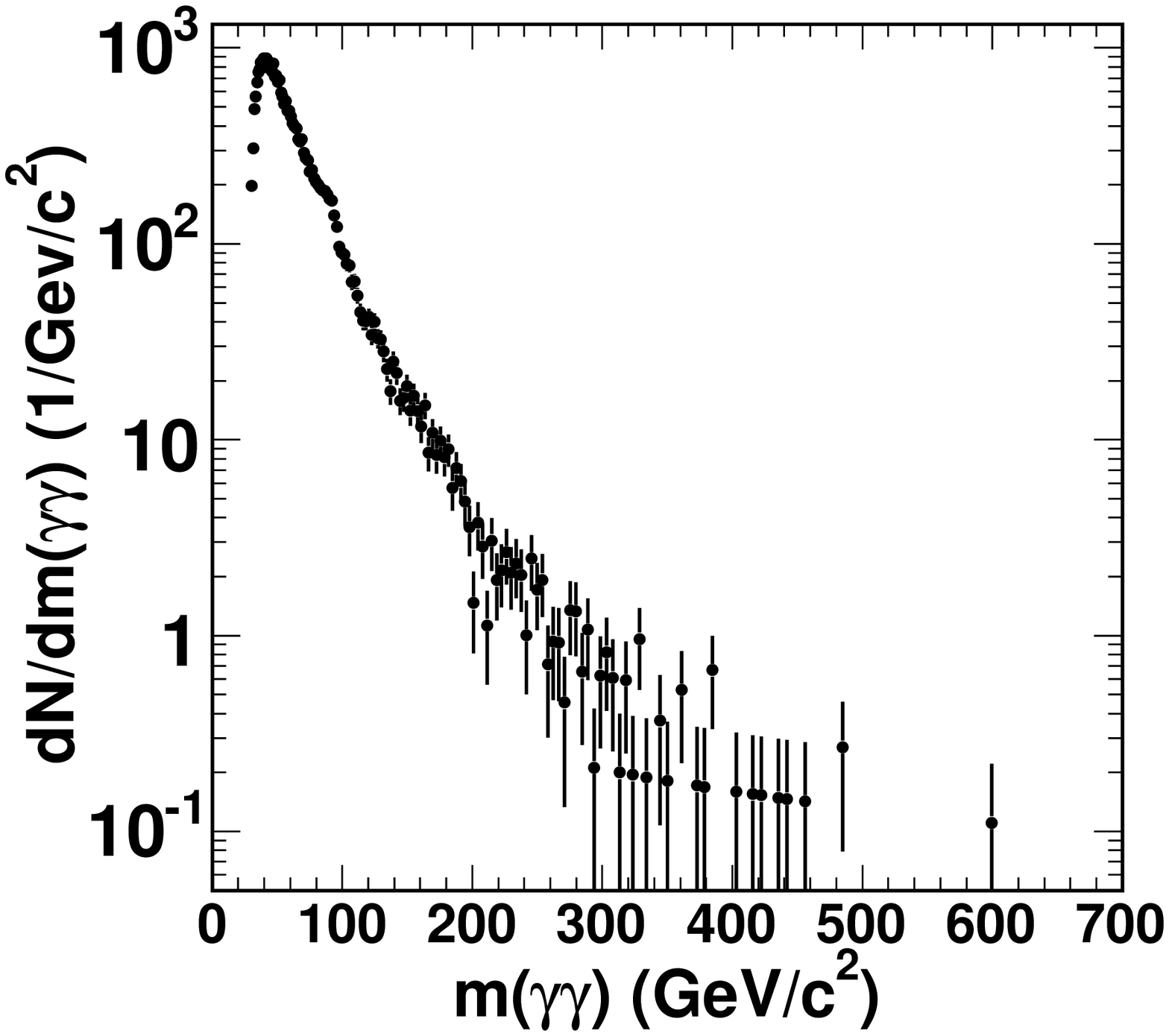}
\end{center}
\caption{The diphoton invariant mass distribution of events for both
CC and CP channels, histogrammed in bins
of approximately one unit of calorimeter mass resolution.
The enhancement near 90~GeV$/c^2$ is due to misidentified $Z$ boson
decays to electrons in the CP sample.}
\label{Fig:all_mass}
\end{figure}

\begin{figure}[h]
\begin{center}
\includegraphics[width=1.0\linewidth]{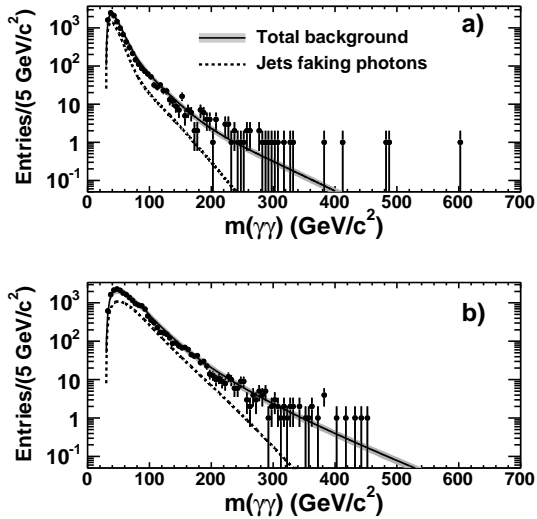}
\end{center}
\caption{
The mass distribution in the CC (a) and CP (b) signal regions
with the {\it a priori} background overlaid. The points are the data. 
The dotted line shows the jets which fake photons 
as predicted from the photon-like jet sample, and the solid line shows
this background plus the {\sc diphox} SM diphoton distribution.
The grey band shows the uncertainty on the total background.
This background is not used in setting limits.}
\label{Fig:display_signal}
\end{figure}

The selected data consist of 11, 088 CC and 20, 933 
CP events. The diphoton invariant mass 
distribution for these events, 
histogrammed in bins equivalent to the mass resolution 
(approximated by $0.13\sqrt{m ({\rm GeV}/c^{\rm 2})}\oplus 0.014m$~GeV, where $m$
is diphoton mass)
is shown in Fig.~\ref{Fig:all_mass}. 
The highest mass pairs occur at 
602~GeV$/c^{2}$ for the CC data and 
454~GeV$/c^{2}$ for the CP data, as shown in 
Fig.~\ref{Fig:display_signal}.

The expected numbers of RS graviton events, as a function of graviton 
mass, are estimated using the {\sc herwig 6.510}
event generator~\cite{Herwig}, with {\sc cteq5l} parton distribution 
functions (PDFs)~\cite{CTEQ5L},
and processed by the {\sc geant~3} based CDF II detector simulation~\cite{Geant}.
The combined acceptance and selection efficiency 
for RS events increases 
from $0.31 \pm 0.01_{stat}$ for gravitons 
of mass 200~GeV$/c^{2}$ 
to $0.36 \pm 0.01_{stat}$ for gravitons of mass 1050~GeV$/c^{2}$.
Final corrections to the efficiency are derived 
by comparing the measured and Monte Carlo simulated detector response to electrons 
from $Z$ boson decays, since a pure sample of reconstructed 
photons is not available, and the characteristics of energy deposited 
in the calorimeter by electrons are 
almost indistinguishable from those of photons.
The largest systematic uncertainties on the expected number of gravitons
arise from the luminosity measurement (6\%) and the choice of PDF (4\%). 
The latter uncertainty is determined from the variation in 
the efficiency when employing different PDF parameterizations.

Most $Z$ bosons that decay to $e^+e^-$ 
are rejected by requiring no associated tracks;
however, approximately 1\% of these remain in the CP sample at a  
mass peak below the search region.
There are two other significant components to the diphoton data sample. 
The first is SM diphoton production, 
which is estimated with the {\sc diphox} next-to-leading-order (NLO) 
Monte Carlo~\cite{diphox} calculation. 
This program computes a cross section as a function of mass. 
The mass spectrum is then corrected by an efficiency function derived 
from a SM diphoton sample generated by {\sc pythia}~\cite{Pythia} 
and processed through the full detector simulation. 
The leading systematic uncertainty on this background estimate 
is approximately 20\%, which comes from the variation in the
cross section when the renormalization scale, $Q^2$, is varied
in the generation.
The second component is jets, 
consisting of quarks that fragment into a high momentum
$\pi^{0}$, which subsequently decays as 
\mbox{$\pi^{0} \rightarrow\gamma\gamma$}.  
The resulting photon showers may overlap, 
and can pass the photon selection. 
The diphoton mass distribution of this background 
is derived from a sample of photon-like jets 
obtained by loosening the photon selection criteria,
then removing the 
events which pass all the signal selection requirements.
 The selection of the photon-like jets is varied and
the resulting variation in the shape of the mass distribution
is approximately 20\%. This is the leading systematic uncertainty 
on the jet background.

To extrapolate to the highest masses, 
the mass distributions of the two background sources 
are fitted to a product of a polynomial and the sum 
of two (for photon-like jets) or three (for {\sc diphox}) exponentials. 
The fit to the photon-like jets includes 
a contribution from the {\sc diphox} shape to allow 
for true diphoton events which appear in this sample.
For the signal region background estimate,
the {\sc diphox} distribution is normalized to the NLO
cross section predicted by {\sc diphox}. 
The jet background shape is normalized to the observed shape 
in the invariant mass region 30~GeV/$c^{2}$ to 100~GeV$/c^{2}$ after 
subtracting the estimated background from SM diphoton production. 

Figure~\ref{Fig:display_signal} shows the observed mass spectra
compared to the predictions. This estimate of the background, 
which shows good agreement with the observation, is only for 
an {\it a priori} comparison, and is not used in 
setting upper limits on a specific production model.

\par To find the most accurate description of the background 
for setting limits, we fit the
CC and CP mass spectra to a polynomial
multiplied by a sum of three components: two exponentials 
and the {\sc diphox} shape.  The normalization of each 
component is allowed to float in the fit.
This function is used for the background estimate as a function of mass 
with statistical uncertainties propagated from the fits.
{\sc diphox} systematic uncertainties, as described above, 
are also propagated.

The background shape is combined with the simulated signal shape 
in a binned likelihood method, with the contents of the
bins treated with Poisson statistics. The 
likelihood as a function of cross section is given by

\begin{equation}
L(\sigma) =\prod^{N_{bins}}_{i=1}\frac{\mu_{i}(\sigma)^{N^{d}_{i}}e^{-\mu_{i}(\sigma)}}{N^{d}_{i}!}~,
\label{eq:likelihood}
\end{equation}

\noindent where \begin{equation} \mu_{i}(\sigma) = \frac{\epsilon \cal{L} \sigma 
\textit{N}^{\textit{s}}_{\textit{i}}}{\textit{N}^{\textit{s}}_{\textit{tot}}} 
+ 
\textit{N}^{\textit{b}}_{\textit{i}}~. \\
\label{eq:mu} 
\end{equation}

Here $N^{d}_{i}$, $N^{b}_{i}$ and  $N^{s}_{i}$ are respectively 
the numbers of observed, background and signal events in bin $i$, and
$N^{s}_{tot}$ is the total number of 
signal events passing selection requirements.
The parameter $\epsilon$ is the total acceptance and efficiency 
for selecting the events predicted by the model, 
$\cal{L}$ is the integrated luminosity, and $\sigma$ is the 
model cross section. Systematic uncertainties 
are included in the likelihood with a gaussian Bayesian prior.

The likelihood function is calculated separately for the 
CC and CP channels, then 
a combined likelihood is formed from the product of the 
two, accounting for the correlations among systematic uncertainties. 
We obtain a posterior density in $\sigma$ assuming a uniform
prior in $\sigma$ applying Bayes' Theorem. The 95\% C.L. upper
limit corresponds to the point below which 95\% of the 
integral of the posterior density lies.

\begin{figure}[h]
\begin{center}
\includegraphics[angle=0,width=1.0\linewidth]{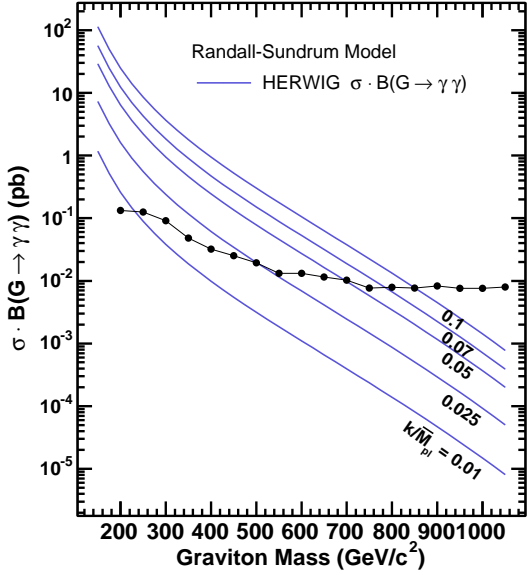}
\end{center}
\caption{The 95\% C.L. upper limit on the production cross section 
multiplied by branching fraction of an RS model graviton decaying to 
diphotons $(\sigma \cdot \mathcal{B}(G\rightarrow \gamma\gamma))$, as a 
function of graviton mass.
Also shown are the predicted $(\sigma \cdot \mathcal{B})$ 
curves for $k/\overline{M}_{Pl}$ = 0.01, 0.07, 0.05, 0.025 and 0.1.}
\label{Fig:upper_limit}
\end{figure}

\begin{figure}[h]
\begin{center}
\includegraphics[width=1.0\linewidth]{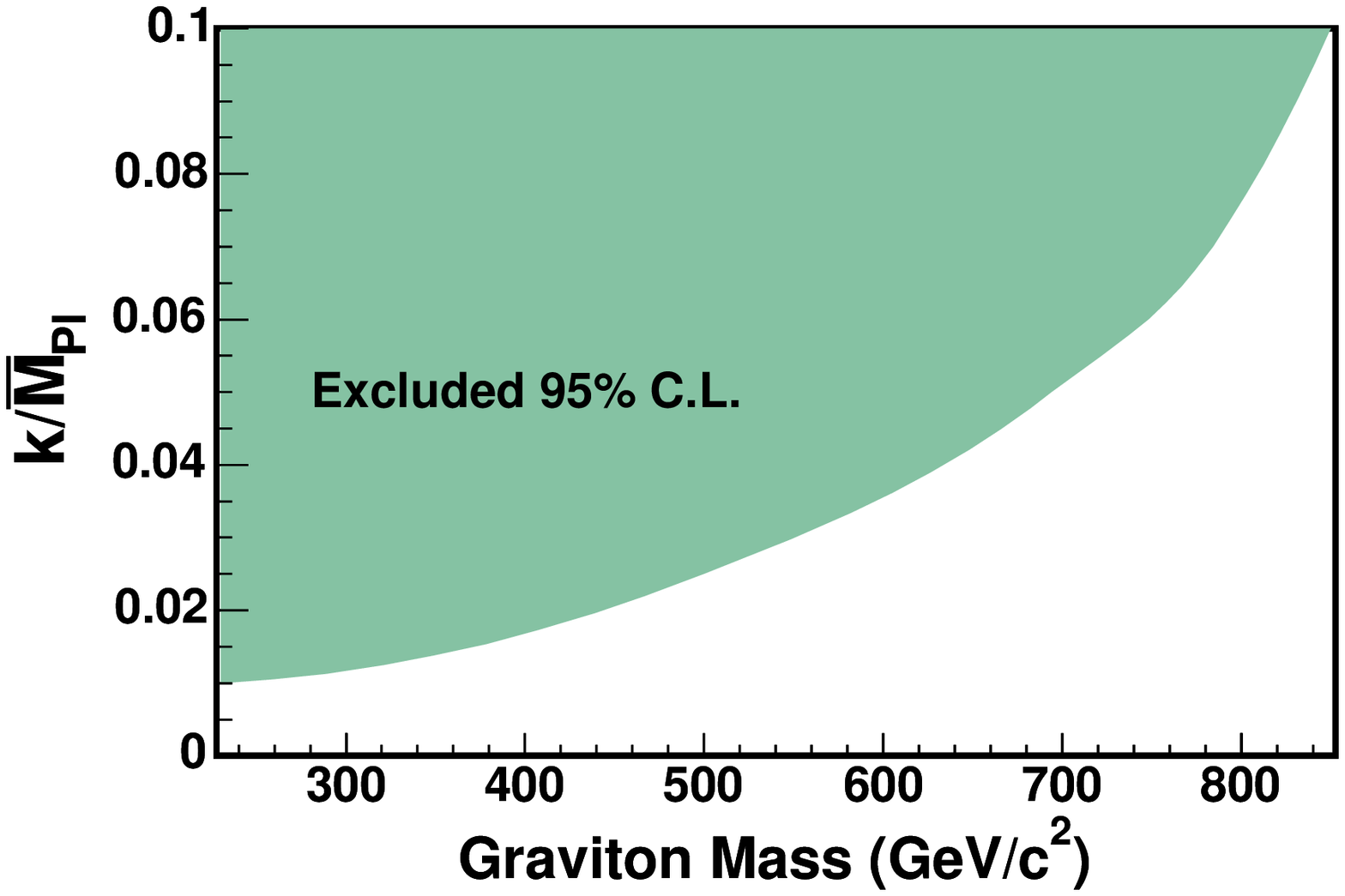}
\end{center}
\caption{The 95\% C.L. excluded region 
in the plane of $k/\overline{M}_{Pl}$ and graviton mass. 
}
\label{Fig:exclude_region}
\end{figure}

The result is shown in Fig.~\ref{Fig:upper_limit}, 
as a function of graviton mass, along 
with the theoretical cross section times branching ratio for RS 
gravitons with $k/\overline{M}_{Pl}$ set to 0.1, 0.07, 0.05, 0.025 and 
0.01.
The leading-order graviton production cross section is multiplied 
by a factor of 1.3, to correct for diagrams at higher-order in 
$\alpha_{s}$~\cite{kfactor}.
From the limit on $\sigma \cdot \mathcal{B}(G \rightarrow \gamma\gamma)$,
lower mass bounds are obtained for the first excited state of 
the RS graviton as a function of the parameter 
$k/\overline{M}_{Pl}$.
The 95\% C.L. excluded region in the $k/\overline{M}_{Pl}$ and graviton 
mass plane is displayed in 
Fig.~\ref{Fig:exclude_region}, with the mass limits summarized 
in Table~\ref{Tbl:mass_limits}.

\begin{center}
\begin{table}
\begin{center}
\begin{tabular}{c c} \hline
$k/\overline{M}_{Pl}$ & Lower Mass Limit (GeV$/c^2$)\\ \hline \hline
0.1 & 850\\
0.07 & 784\\
0.05 & 694\\
0.025 & 500\\
0.01 & 230\\ \hline
\end{tabular}
\caption{The 95\% C.L. lower limits on the mass of the RS 
graviton for the specified values of $k/\overline{M}_{Pl}$.}
\label{Tbl:mass_limits}
\end{center}
\end{table}
\end{center}

In conclusion, we have searched for evidence of an anomalous peak in the 
diphoton mass spectrum using approximately 1.2~fb$^{-1}$ of data
collected by the CDF II Detector at the Fermilab Tevatron.
Our sample includes events with both photons in the central detector, 
and events with one photon in the central detector 
and one in the plug detector.
We find no evidence of new physics. We evaluate one model
of hypothetical new diphoton production and exclude 
RS gravitons below masses ranging 
from 230~GeV/$c^{2}$ to 850~GeV$/c^{2}$,
for a coupling parameter $k/\overline{M}_{Pl}$ of 0.01 to 0.1, at the 95\% C.L.
This results in a significant improvement, at high mass, over the previous best 
available limit.

We thank the Fermilab staff and the technical staffs of the participating 
institutions for their vital contributions. This work was supported by 
the U.S. Department of Energy and National Science Foundation; the 
Italian Istituto Nazionale di Fisica Nucleare; the Ministry of Education, 
Culture, Sports, Science and Technology of Japan; the Natural Sciences and 
Engineering Research Council of Canada; the National Science Council of the 
Republic of China; the Swiss National Science Foundation; the A.P. 
Sloan Foundation; the Bundesministerium f\"ur Bildung und Forschung, Germany; 
the Korean Science and Engineering Foundation and the Korean Research 
Foundation; the Science and Technology Facilities Council and the 
Royal Society, UK; the Institut National de Physique Nucleaire et 
Physique des Particules/CNRS; the Russian Foundation for Basic Research; 
the Comisi\'on Interministerial de Ciencia y Tecnolog\'{\i}a, Spain; 
the European Community's Human Potential Programme; the Slovak R\&D Agency; 
and the Academy of Finland. 

\begin{thebibliography}{99.}
\bibitem{CDF} D. Acosta {\it et al.}, Phys. Rev. D {\bf 71}, 032001 (2005).
CDF uses a cylindrical coordinate system in which $\phi$ 
is the azimuthal angle, $r$ is the radius from the nominal
beamline, and $+z$ points along the direction of the proton beam.
The pseudorapidity is defined as $\eta = -\ln[\tan(\theta/2)]$,
where $\theta$ is the polar angle measured from the $+z$ axis.
Calorimeter energy (track momentum) measured in the
plane transverse to the beam is denoted as $\etm$ ($\ptm$).
\end{thebibliography}
\end{document}